\documentclass[reprint,aps,prd,twocolumn,superscriptaddress,nofootinbib,showkeys]{revtex4-2}
\usepackage[utf8]{inputenc}
\usepackage[colorlinks=true, linkcolor=blue, citecolor=blue, urlcolor=blue]{hyperref}
\usepackage{amsmath, amssymb, amsfonts, bm, graphicx, microtype, orcidlink, esvect, url, booktabs, amsthm, adjustbox, placeins, csquotes, changes, xspace, multirow, bm}
\usepackage[normalem]{ulem}

\newcommand{\m}{M\textsubscript{\(\odot\)}\xspace}
\usepackage{xcolor}
\definecolor{kkcol}{HTML}{bc82bf}

\begin{document}

\title{Search for Sub-Solar Mass Binaries in the First and Second Part of LIGO’s Fourth Observing Run O4a-O4b}

\author{Keisi Kacanja\orcidlink{0009-0004-9167-7769}}
\email[Corresponding author: ]{kkacanja@syr.edu}
\author{Kanchan Soni\orcidlink{0000-0001-8051-7883}}
\author{Alexander H. Nitz\orcidlink{0000-0002-1850-4587}}
\affiliation{Department of Physics, Syracuse University, Crouse Dr, Syracuse, NY 13210}

\date{\today} 

\begin{abstract} 
The observation of sub-solar mass neutron stars and black holes would signify a new class of astrophysical objects since these systems are not expected to form via standard stellar evolution. We present the first sub-solar mass search for binaries consisting of neutron star or black holes that uses the data from both the second part (O4b) and first part (O4a) of the fourth observing run of the Advanced LIGO detectors. The search covers primary masses 0.1 to 2 $M_\odot$ and secondary masses 0.1 to 1 $M_\odot$. Tidal effects up to $7\times10^5$ are accounted for in extremely low-mass neutron stars. No statistically significant candidates are identified. We place a $90\%$ confidence upper limit on the merger rate $\mathcal{R}_{90}$ for sub-solar mass black holes and neutron stars to be $< 5.2\times10^{2}\,\textrm{Gpc}^{-3}\,\textrm{yr}^{-1}$ for a chirp mass of 0.5 $M_\odot$,  a factor $2.2 \times$ lower compared to previous O1-O4a searches. We constrain the effective local dark matter fraction to be  $\tilde{f}_\textrm{PBH}<0.6\%$ for $M_{\textrm{PBH}}=0.8 \,M_\odot$, approximately $1.5$ factor improvement to previous O1-O4a constraints. Given our model assumptions, our local dark matter fraction constraints are $2-7$ times lower than the OGLE microlensing survey for $M_{\textrm{PBH}}\ge{0.6}~M_\odot$.

\end{abstract}


\maketitle

\section{Introduction} \label{sec:intro}

The era of gravitational-wave (GW) astronomy began in 2015 with the Advanced LIGO detectors, culminating in the first direct detection of GW150914, from a binary black hole (BBH) merger \citep{advancedligo,2016PhRvL.116f1102A}. This was followed by the first binary neutron star (BNS) observation, GW170817, which was accompanied by electromagnetic counterparts across the spectrum \citep{2017PhRvL.119p1101A,Alexander:2017aly,2021ApJ...918L..29R,2017Sci...358.1556C,2017ApJ...848L..30P,2019ApJ...887L..24M,2019ApJ...887L..21R,LIGOScientific:2017ync,Valenti_2017}, and the subsequent detection of neutron star black hole (NSBH) mergers \citep{2021ApJ...915L...5A}. Across the completed observing runs, the LIGO-Virgo-KAGRA (LVK) network has reported hundreds of GW events, with the vast majority BBH mergers, together with a small number of BNS and NSBH systems \citep{1gwtc,2gwtc,3gwtc,4gwtc}. These observations have largely been consistent with expectations from stellar evolution, with component masses of the binary above $1 M_\odot$ \citep{4gwtcpop}. 

To date, no confirmed GW detections of sub-solar mass (SSM) compact binaries have been made. This absence may be due to the current limitations of detector sensitivities or the intrinsic rarity of these systems. However, there has been one optical observation of the supernova remnant HESS J1731-347, which suggests the central compact object may be a very light neutron star with an estimated mass of $0.77\m$ and a radius of 10 km \citep{2022NatAs...6.1444D}. Interpreting this event as a SSM neutron star is challenging, as mass and radius constraints from GW170817 disagree with the constraints provided by this isolated binary. Alternative studies have evaluated the event against other unique formation scenarios and showed consistency with strange stars or hybrid star model representations of the event \citep{Horvath:2023uwl,Mariani:2024gqi}.

The theoretical minimum stable neutron star mass is predicted to be approximately $0.1~\m$ \citep{1983bhwd.book.....S}; however, standard stellar evolution is not expected to produce such remnants. One proposed formation mechanism involves the fragmentation of neutrino-cooled accretion disks surrounding rapidly rotating collapsing stars \citep{Piro:2006ja,Metzger_2024,Chen:2025uwd,Lerner:2025dkd}. Hydrodynamic simulations have shown that under sufficiently high cooling rates, the disk can produce multiple fragments of neutron stars ranging from $0.01-1~\m$ in binaries and potentially be observable with GWs \citep{Chen:2025uwd}. However, interaction of the fragments with the disk would produce GW that are very distinct from isolated binary models and may be missed in traditional searches \citep{Lerner:2025dkd}.

GW searches provide a unique opportunity to observe a potential SSM population. Unlike electromagnetic observations, GW signals from dense star systems encode information about the dense matter equation of state (EOS) through tidal interactions during the inspiral \citep{2007Ap&SS.308..371L,2021ApJ...918L..29R,Rutherford:2024srk,Raaijmakers_2021}. As neutron stars orbit one another, deformations due to the tidal field from the companion leave measurable imprints on the waveform, quantified by the effective tidal deformability ($\tilde{\Lambda}$) \citep{2008PhRvD..77b1502F,2010PhRvD..81l3016H}. Since lighter neutron stars are less compact, they are expected to exhibit substantially larger tidal deformabilities ($\Lambda$) than their heavier counterparts. EOS models predict $\Lambda$ ranging from $\mathcal{O}(10^3)$ for a $1~\m$ neutron star to as large as $\mathcal{O}(10^8)$ for a $0.1~\m$ neutron star \citep{2020CQGra..37d5006A,1997A&A...328..274B,MULLER1996508,DANIELEWICZ200936,PhysRevC.72.014310,PhysRevC.53.740,Capano:2019eae}.

The detection of an SSM BNS would provide a particularly sensitive probe of the EOS \citep{2024arXiv240517326K}. Observations by the Neutron Star Interior Composition Explorer (NICER), together with GW170817, have significantly improved constraints on the neutron star EOS near the canonical neutron star mass range \citep{2019ApJ...887L..21R,2019ApJ...887L..24M}. However, these measurements are considerably weaker in the SSM range because no such binary systems have yet been observed. GW searches targeting SSM binaries provide a unique opportunity to extend EOS constraints into an unexplored mass range in the case of an observation.

In the absence of measurable tidal effects in the signal, neutron star interpretations would be disfavored, and the consideration of black holes would be motivated. Standard stellar evolution is not expected to produce SSM black holes, and such observations would point toward alternative formation channels. Primordial black holes (PBHs), which are hypothesized to form in the early Universe through the collapse of overdense regions, provide one possibility for forming SSM BBHs \citep{2024PhRvD.109l4063C,PBHreview}. PBHs can have masses ranging anywhere from $\sim 10^{-16}- 10^2~\m$ depending on their formation epoch \citep{greenPBHreview}. PBHs with masses $0.1-1~\m$ are predicted to have formed during the cosmic QCD epoch \citep{Byrnes:2018clq, Clesse:2020ghq, Jedamzik:2020omx, Jedamzik:2020ypm, Juan:2022mir}. Additionally, if the observed SSM BBH is not formed through primordial channels, alternative exotic channels, such as black holes formed through dark matter (DM) induced collapse or dark sector compact objects, may explain such events \citep{Shandera:2018xkn,Dasgupta:2020mqg,Singh:2020wiq,Oguri:2017ock,Croon:2020ouk}.

The nature of DM remains an open question. Current theories predict that DM may consist of subatomic particles, such as WIMPs or axions. It could also be composed of macroscopic compact objects such as PBHs \citep{PBHreview,darkmatterreview,PBHasDM}. In this scenario, PBHs themselves could constitute a fraction of the observed DM density. Alternatively, if DM is composed of particles, PBHs formed in the early Universe may retain surrounding DM halos due to their formation in a largely pristine environment, and provide a direct measurement of DM if observed \citep{darkmatterreview}. Interactions between a PBH binary and the surrounding DM distribution could leave detectable imprints on the GW signal observed by ground-based interferometers \citep{2023PhRvD.107h3006C,Boudaud:2021irr}, although such effects remain challenging to observe. Observational constraints from microlensing, stellar dynamics, and cosmic microwave background measurements limit the allowed contribution of PBHs to the DM abundance across much of the parameter space \citep{Macho:2000nvd,EROS-2:2006ryy,Griest:2013aaa,Niikura:2019kqi,Mroz:2024mse,PBHbounds}. In the SSM regime, GW searches provide complementary constraints by limiting the merger rate of SSM PBH binaries \citep{greenPBHreview,ligosubsolar,nitzsubsolar,nitzsubsolar1,nitzsubsolar2,nitzsubsolarhighq}.

During the third part of the fourth observing run (O4c), low-significance GW alerts have identified potential SSM candidates. On August 18, 2025, S250818k was detected with a false alarm of 2.1 $\text{yr}^{-1}$, with at least one component expected in the SSM range \citep{GraceDB_S250818k}. Temporal and spatial coincidences of S250818k initially corresponded with the optical transient ZTF25abjmnps (AT2025ulz) \citep{S250818k,ZTF25abjmnps}, providing a potential multi-messenger association. Subsequent follow-up study classified AT2025ulz as a Type IIb supernova \citep{Gillanders:2025fwf}. The second low-significance GW alert occurred on November 12, 2025, where S251112cm was estimated to have a false alarm rate of 1 per 6.2 years \citep{S251112cmgcn,Vieira:2026eof,Haque:2026yum}. Several followup studies were performed to search for a potential optical counterpart, but no convincing candidate has been identified \citep{Vieira:2026eof,Hall:2026gov}. Although there is no associated optical counterpart, this may be consistent as a PBH scenario, which remains viable within current observational constraints \citep{Haque:2026yum}. However, uncertainties in the PBH abundance, merger rate, and the significance of the GW candidate prevent a definitive interpretation, and a more robust event will be necessary \citep{Haque:2026yum}. If both these events are astrophysical, it could provide the first tentative evidence for SSM mergers, potential observation of PBHs, or the observation of objects formed through fragmentation scenarios.

Multiple searches and followup analyses targeting SSM binaries have been performed using data ranging from the O1 to O4b observing runs \citep{nitzsubsolar,nitzsubsolar1,nitzsubsolar2,ligosubsolar,Kacanja:2026byy,Nitz:2021zwj,lvkssmcontinuous,lvko4aSSM,ssmcontinuous,2025ApJ...984...61K,Yuan:2024yyo,o4ape}. These include targeted searches for high mass-ratio SSM binaries \citep{nitzsubsolarhighq} as well as searches for planetary-mass compact objects \citep{lvkssmcontinuous,ssmcontinuous}. However, most previous analyses have neglected the effects of tidal deformation, limiting their sensitivity to neutron star-like systems. This omission can significantly reduce signal recovery for low-mass binaries, with losses of up to $78.4\%$ for an equal-mass $0.2~\m$ binary \citep{2023PhRvD.107j3012B}. In practice, these estimates are optimistic, as additional sensitivity losses arise from search techniques used to mitigate the effects of non-Gaussian detector noise \citep{Usman:2015kfa,2017PhRvD..95d2001M}.

To address this limitation, we previously developed a preliminary modeled SSM search incorporating tidal deformability effects up to $\tilde{\Lambda}\sim10^{4}$ using O3 data \citep{2025ApJ...984...61K}. However, this range does not fully encompass the extremely large tidal deformabilities expected for very low-mass neutron stars. For example, neutron stars with masses near $0.1~\m$ may exhibit tidal deformabilities as large as $\sim10^{7}-10^{8}$ depending on the EOS \citep{Capano:2019eae}. In our subsequent O4a search, we expanded the tidal deformability range to include more physically motivated neutron stars, informed by a softer EOS. In this search, we use the most plausible EOS constraints consistent with GW170817 from \citet{Capano:2019eae} to define the maximum tidal deformability. This expansion increases the search space by $24\%$ relative to the previous O4a analysis and by $150\%$ relative to our preliminary O3 SSM search.

We present the first SSM search using the O4a+O4b observing runs that is sensitive to both BBH and BNS-like systems, including plausible EOSs for sources down to $\sim0.25~\m$. We reanalyze the O4a data with the newly expanded search space from the preliminary O4a search. This combined O4a+O4b search is enabled by efficient ratio-filter de-chirping methods that allow for computationally tractable exploration of the expanded parameter space \citep{fir,2025arXiv251112894M}. We identify no significant SSM candidates with a false alarm rate $\le 0.01~\text{yr}^{-1}$ and place improved constraints on the merger rates of these systems, obtaining $\mathcal{R}_{90}=1.1\times10^{5}$ for $\mathcal{M}=0.1~\m$ and $\mathcal{R}_{90}=1.5\times10^{2}\,\textrm{Gpc}^{-3}\,\textrm{yr}^{-1}$ for $\mathcal{M}=1.1~\m$. 

\section{Search} \label{sec:population}
We perform a matched filtering analysis using the \textsc{PyCBC} framework \citep{Nitz:2017svb,Usman:2015kfa}. In this approach, the detector data is correlated with a template waveform to measure the signal-to-noise ratio (SNR). The SNR quantifies the strength of the similarity between the modeled waveform and the data. The peak of the SNR time series beyond some threshold is extracted and is often referred to as a trigger. The triggers are analyzed for time and parameter coincidences across the detector network. A ranking statistic is assigned to the triggers to help distinguish astrophysical candidates from noise fluctuations \citep{2017ApJ...849..118N}. Coincident triggers are then assigned statistical significance by comparing to the empirical distribution measured using time slides. Candidates exhibiting a large ranking statistic and a high network SNR, and correspondingly low false alarm rate (FAR), are considered the most credible detections \citep{LIGOScientific:2016aoc}. 

\begin{figure*}[htbp]
    \centering
    \begin{tabular}{cc}
        \includegraphics[width=0.48\textwidth]{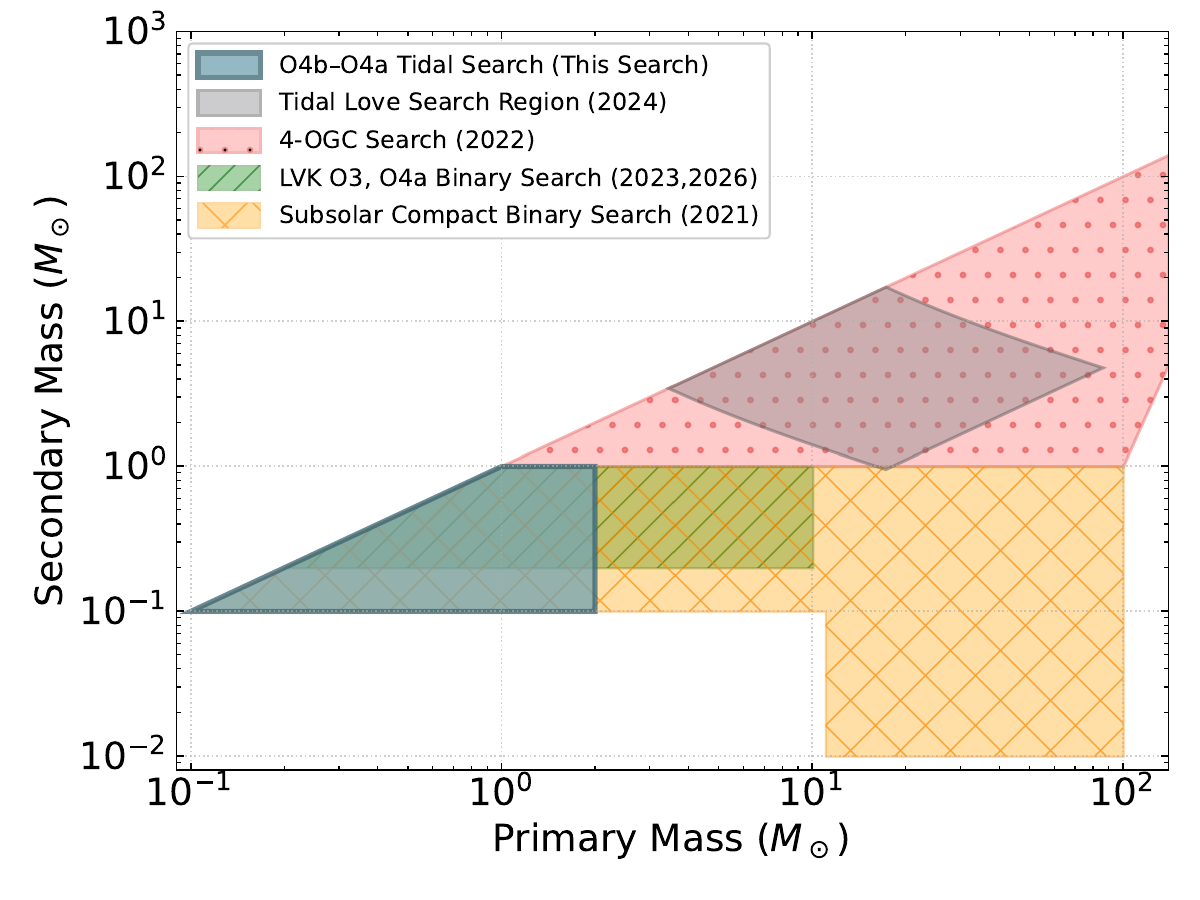} 
        & 
        \includegraphics[width=0.48\textwidth]{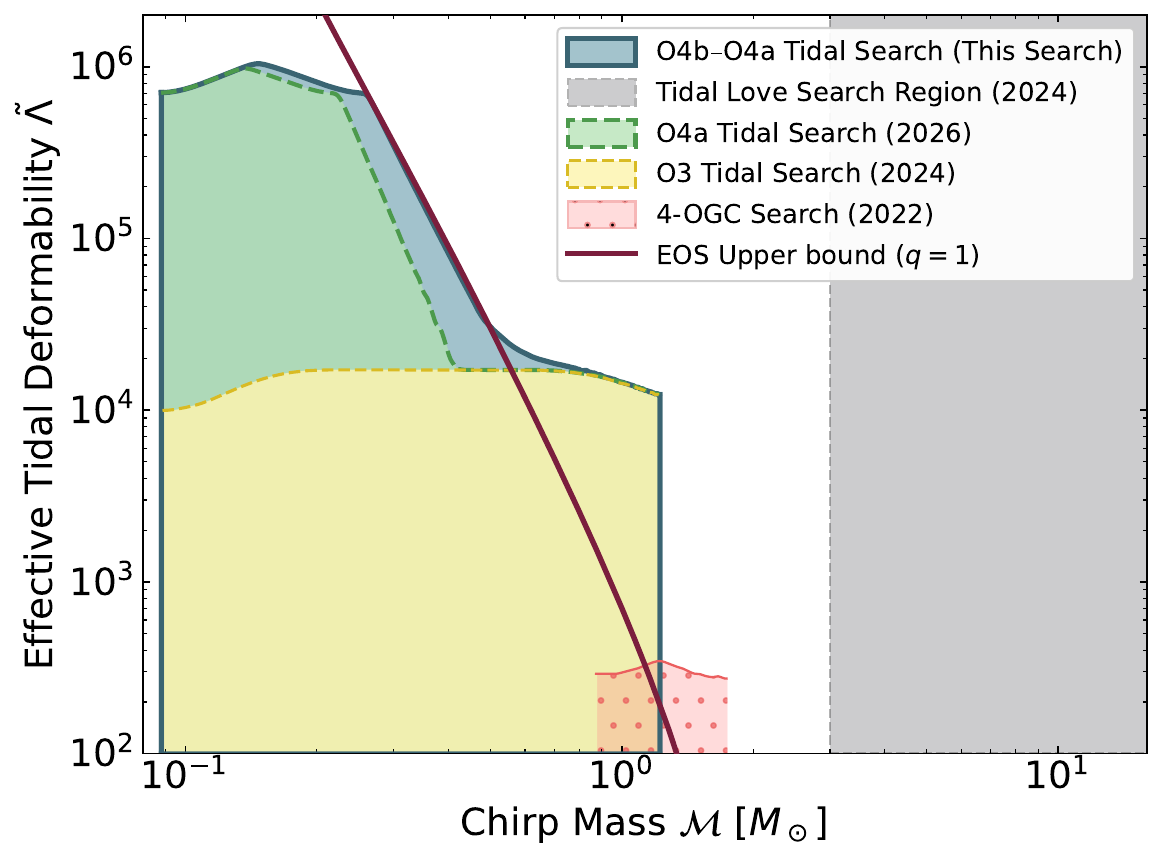}
    \end{tabular}

    \caption{SSM and Tidal Searches.
    \textbf{Left Panel:} Component-mass parameter space covered by various searches, including the present O4a+O4b search (teal), LVK SSM searches (green) \citep{ligosubsolar,lvko4aSSM}, 4-OGC (dotted pink) \citep{Nitz:2021zwj}, and the broad SSM search (hashed yellow) \citep{nitzsubsolar}, and tidal searches (gray) \citep{Chia:2023tle}.
    \textbf{Right Panel:} Previous tidal searches as a function of chirp mass $\mathcal{M}$, and effective tidal deformability $\tilde{\Lambda}$. The teal region shows the parameter space explored in the present O4a+O4b search, while the pastel yellow and mint green regions show the previous O3 \citep{2025ApJ...984...61K} and O4a \citep{Kacanja:2026byy} tidal searches, respectively. The gray region shows the parameter space covered by the previous tidal Love-number search \citep{Chia:2023tle}. The dark red line corresponds to the EOS upper limit bounds most consistent with GW170817 \citep{Capano:2019eae} converted to $\mathcal{M}$ and $\tilde{\Lambda}$ space using a mass ratio $q=1$.
    }
    \label{fig:combined_banks}
\end{figure*}

Since the properties of a true GW signal is not known in advance, banks of templates are constructed to densely cover the parameter space and ensure minimal loss in the recovered SNR. We utilize stochastic placement algorithms to incorporate tidal effects characteristic of SSM neutron stars into the template bank \citep{2024ApJ...975..212K,2021PhRvD.104d3008H,2009PhRvD..80j4014H}. The search space explored in this work has been previously developed in our preliminary O3 search with $m_1\in[0.1,2]~\m$, $m_2\in[0.1,1]~\m$, and $\lambda_{1,2} \leq 10^{4}$ \citep{2025ApJ...984...61K}, and O4a search \citep{Kacanja:2026byy}, where the tidal space is informed by a softer EOS. In this search, we cover the same masses, but utilize a more realistic EOS model consistent with GW170817 from \citet{Capano:2019eae}, shown in dark red in the right panel of Figure \ref{fig:combined_banks}. A full exploration of the EOS tidal limits would yield an unfeasible large bank on the order of $\sim 10^{8}$ templates; thus we place an additional cut on $\lambda_{1,2} \leq 7\times10^{5}$. Since PBHs and SSM BNSs are not expected to be fast spinning, we impose our effective spin \citep{Ng:2018neg} bounds to be small $\chi_{\textrm{eff}} \in [-0.05, 0.05]$ \citep{Mirbabayi:2019uph,DeLuca:2019buf,Franciolini:2022iaa,Chu:2024rwt,2017ApJ...846..170T}. 

We model isolated binary systems using \textsc{TaylorF2} approximant at 3.5 Post-Newtonian order \citep{Droz:1999qx,Blanchet:2002av,Faye:2012we}. This choice does not significantly impact the recovered SNR. This was verified by calculating the fitting factors \citep{Buonanno:2002fy} between \textsc{IMRPhenomXP\_NRTidalv2} \citep{Colleoni:2023ple} injections with the \textsc{TaylorF2} template bank, finding consistent recovered SNRs across the parameter space considered. Since these signals are extremely long in duration and are expensive to search for, we impose a upper frequency cutoff of 800 Hz, a global lower frequency cutoff of 20 Hz, and a maximum signal duration of 512 seconds by dynamically adjusting the lower frequency cutoff \citep{nitzsubsolar}. For the lowest-mass binaries, with $m_{1,2}=0.1~\m$, this results in a lower frequency cutoff of 67.2 Hz ($3\%$ of the bank). Since tidal effects are most pronounced at higher frequencies, we verified that high frequency cuts do not lead to a measurable loss in recovered SNR ($\lesssim 0.01\%$ throughout the bank space). The final bank consists of 31 million templates.

This search is made feasible by recent advances in matched-filtering techniques \citep{fir,2025arXiv251112894M}, which use compact finite-impulse-response (FIR) representations of waveform models. We achieved an eightfold improvement in template-processing speed per core on a high-contention cluster in comparison to the standard matched filtering. The FIR representation is constructed using the ratio between a target waveform and a reference waveform. The reference bank was constructed with a minimal match of 0.6 and was used to tile the target bank (constructed with a minimal match of 0.95). Because waveform generation is required only for the reference templates, the computational cost can be further reduced by pre-generating the reference waveforms. A low order spline interpolant is constructed for each reference waveform. The spline nodes are proposed dynamically until the interpolation of the points reconstructs the original waveform with a mismatch of less than $<0.0001\%$. The points are saved and used for re-interpolation during matched filtering, skipping the need to generate the full waveform. This method provides an additional $\sim20-30\%$ reduction in computational cost.

\begin{figure}
    \centering
    \includegraphics[width=\linewidth]{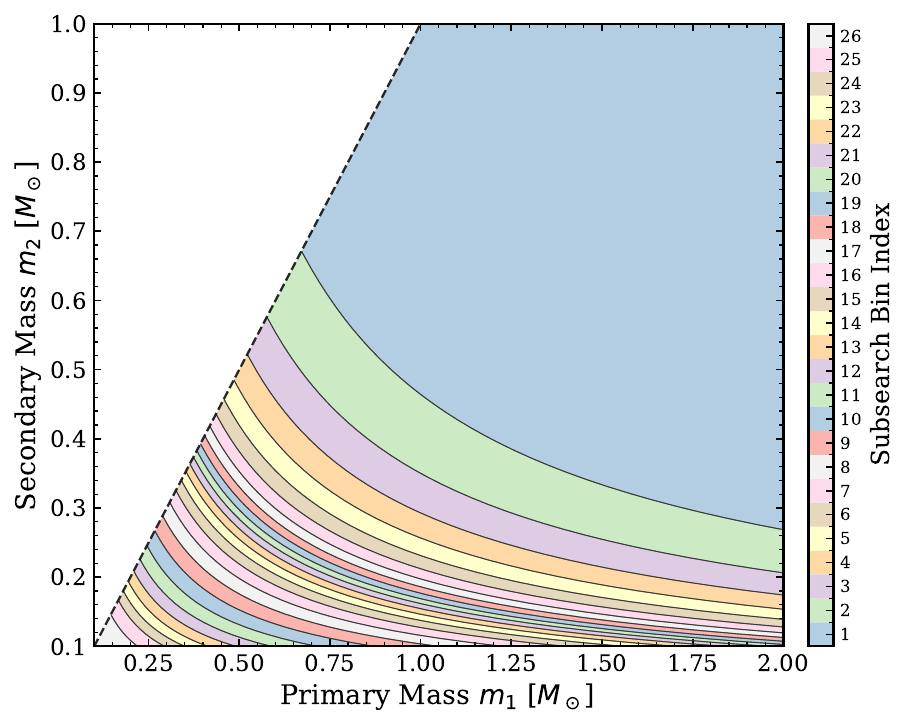}
    \caption{The overall search is separated into 26 sub-searches shown in the various colors, with each bin covering $\sim1.2$ million templates.}
    \label{fig:sub26}
\end{figure}

The final template bank is shown in Figure~\ref{fig:combined_banks} in teal alongside those used in previous SSM searches. Previous modeled SSM searches did not incorporate tidal deformability and therefore used BBH-like waveform models. The LVK SSM searches covered $m_1\in[0.2,10]~\m$ and $m_2\in[0.2,1]~\m$, with spins up to 0.1 for $m_{1,2}<0.5~\m$ \citep{ligosubsolar,lvko4aSSM}. Their template bank contained approximately 2 million templates, with a fixed lower frequency of 45 Hz and an upper frequency cutoff of 1000 Hz. The search of \citet{nitzsubsolar} extended the parameter space to lower masses, covering $m_1\in[0.1,7]~\m$ and $m_2\in[0.1,1]~\m$, while also incorporating eccentricity up to $e=0.3$, with approximately 7.8 million templates. 

Our previous O3 search was the first modeled SSM analysis to incorporate tidal deformability, using $\lambda_{1,2}\leq10^4$ \citep{2025ApJ...984...61K} and a bank of roughly 10 million templates. In the O4a tidal search, the upper limit on tidal deformability was chosen using a soft EOS, further increasing the search space to 25 million templates \citep{Kacanja:2026byy}. For this present analysis, the new EOS bounds extend this search to $31$ million templates, an additional 21 million templates from the O3 search, and 6 million from the O4a tidal search. The resulting bank represents one of the largest modeled SSM searches to date, making computationally efficient matched-filtering techniques, such as the FIR representation described above, essential for making the search computationally tractable.

Because of the large size of the template bank, constructing detector coincidences through time slides represents a high computational and memory cost. For example, performing time slides over three days of data required approximately 900 GB of memory in one case, exceeding the available memory on many computing nodes. To make the search computationally tractable, we divide the full parameter space into 26 independent sub-searches, as shown in Figure~\ref{fig:sub26}. The bins are organized in terms of chirp time $\tau_0$ \citep{Cokelaer:2007kx}, with each sub-search containing approximately 1.2 million templates, substantially reducing the memory requirements for the search postprocessing steps.

We additionally performed a search for single-detector triggers separately from the coincident search. Because the background estimation depends on the underlying trigger distribution and the amount of background time available, we combined triggers from all analyzed times into one global dataset to ensure a robust background estimation. Since the trigger rates are exceptionally large, we performed the single analysis across 26 different sub-searches. The initial search revealed that Livingston data contained unique non-stationary transient noise that was not adequately removed or suppressed by the standard data-quality checks. We applied a non-stationary time heuristic which vetoes times where the local trigger rate exceeds that expected for astrophysical sources; this is conceptually similar to previously proposed bank vetoes~\cite{Hanna2008, Harry:2010fr}. The resulting cleaned trigger set was used to characterize the single-detector background independently of the coincident search.

\section{Results}

We analyzed data from the publicly available O4 observing run of the Advanced LIGO detectors, consisting of O4a spanning May 24, 2023 (GPS 1368195220) to January 16, 2024 (GPS 1389456018) and O4b spanning April 10, 2024 (GPS 1396796418) to January 28, 2025 (GPS 1422118818) \citep{Trovato:2019liz,GraceDB}. Virgo data was not included in this analysis because the sensitivity during O4 did not provide a significant improvement. Instead, we incorporated single-detector analyses, resulting in a $\sim 20\%$ improvement in sensitivity compared to the coincident-detector search alone. No new significant SSM candidates were identified in the combined O4a+O4b dataset. The most significant candidate has a FAR of 0.6 per year. The list of candidates are shown in Table~\ref{tab:candidates}.

\begin{table}[ht]
    \centering
    \caption{Candidates followed up with parameter estimation, ordered by significance, including intrinsic source parameters, merger time, FAR, network SNR, and the detectors that discovered the trigger. Detector HL represents Hanford and Livingston events and L as Livingston only events The FAR accounts for the trials factor. Parameter estimation was not performed on the single detector triggers. After the followup analysis, the HL candidates are consistent with noise.}
    \label{tab:candidates}
    \begin{adjustbox}{width=\columnwidth}
        \begin{tabular}{lccccc}
            \toprule
            GPS Time & $\mathcal{M}$ [$M_\odot$] & $\tilde{\Lambda}$ & FAR [$\text{yr}^{-1}$] & Network SNR & Detector \\
            \midrule
            1403058723.3752 & 0.306 & $2.35\times10^{5}$ & 0.60 & 9.04 & HL \\
            1414544007.7434 & 0.166 & $2.33\times10^{5}$ & 1.60 & 9.24 & HL \\
            1416931638.2759 & 0.301 & $1.65\times10^{3}$ & 3.60 & 9.22 & HL \\
            1406523988.8291 & 0.666  & $3.00\times10^{0}$ & 4.73 & 6.19 & L  \\    
            1420047316.9424 & 0.388 & $1.15\times10^{4}$ & 5.28 & 9.51 & HL \\
            1416603990.9209 & 0.586 & $1.89\times10^{3}$ & 6.87 & 6.02 & L  \\
            1410580727.5210 & 0.157 & $4.27\times10^{5}$ & 7.96 & 9.32 & HL \\
            \bottomrule
        \end{tabular}
    \end{adjustbox}
\end{table}

We followed up all of the two detector triggers listed in Table \ref{tab:candidates} with a complementary parameter estimation using the \textsc{PyCBC} inference framework \citep{Biwer:2018osg}. For most of these triggers, the elevated search statistics were associated with transient noise. When the data was gated around the transient, the recovered SNRs was broadly consistent with those reported by the search. However, gating can alter the inferred signal properties and may artificially increase the recovered SNR. We therefore repeated the parameter estimation using nearby data segments in regions excluding the identified transient noise. In these cases, the recovered SNRs decreased by approximately $\sim 2$ for most candidates. Overall, the majority of the investigated triggers are consistent with transient noise rather than astrophysical signals. For example, our most statistically significant candidate (1403058723.3752) has transient noise 2 seconds before the trigger time in the Livingston detector. Performing the inference over just the Hanford data reduces the SNR to 5.37, which is not significant.

\section{Results: Constraints on SSM Binary Mergers}

\begin{figure}
    \includegraphics[width=\linewidth]{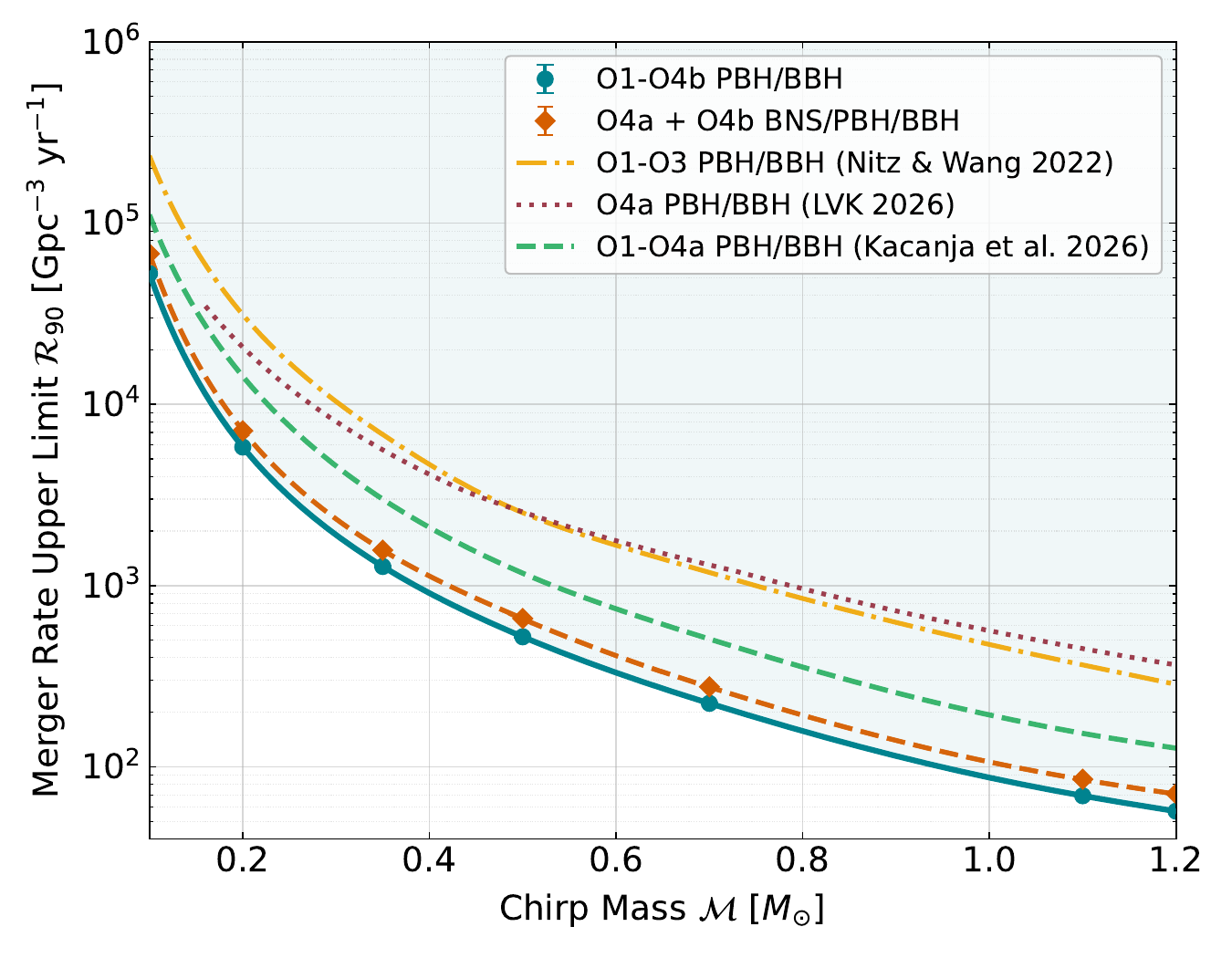}
    \caption{Upper limit constraints on the merger rate at a $90\%$ confidence $\mathcal{R}_{90}$ as a function of chirp mass $\mathcal{M}$. The solid teal curve with circular markers represents the global combined rate from O1 through O4b, incorporating the O1–O3 constraints from \citet{nitzsubsolar} (shown in dashed dotted yellow) combined with the O4a+O4b coincidence and single-detector analyses (shown in dashed orange). For comparison, the latest O4a LVK SSM search constraints, reweighted to a uniform chirp mass prior, are shown in the dark red \cite{lvko4aSSM}. The cumulative O1-O4b has roughly a $2\times$ improvement to the previous O1-O4a rates.}
    \label{fig:merger-rate}
\end{figure} 

We can place 90\% confidence upper limits on the local merger rate by estimating the sensitive volume by constructing fiducial signals and observing the number of found and missed signals from the search pipeline. The number of found injections $N$ can be described as follows

\begin{equation}
     N = \int dt \int d\theta dz  \frac{dV}{dz} \frac{d\mathcal{R}(\theta,z)}{d\theta} \frac{\epsilon(\theta, z)}{(1+z)}\,,  
\end{equation}

where $dV/dz$ represents the differential comoving volume and the differential merger rate density is $d\mathcal{R}(\theta,z)/d(\theta)$. The intrinsic and extrinsic parameters of the binary are represented by $\theta$, the redshift is $z$, and $t$ is the observation time. The detection efficiency $\epsilon(\theta,z)$ is the probability of measuring a detection given $\theta$ and $z$.  Since current GW detectors are sensitive to nearby binaries, we can simplify the following expression as 

\begin{equation}
     N = {R}(\theta_0)\int dtdVd\theta \epsilon(\theta) \,.
\end{equation}
The detection efficiency still depends on redshift. For simplification, redshift is absorbed into $\theta$. We can define ${\langle VT\rangle}_{\theta_0} = \int dtdVd\theta \epsilon(\theta)$. 

Given a null detection, we can calculate an upper limit of the merger rate $\mathcal{R}_{90}$ using the loudest event method \cite{2009CQGra..26q5009B} which states that 

\begin{equation}
     \mathcal{R}_{90}= \frac{2.303}{\langle VT\rangle}\,.
\end{equation}

where 2.303 is the $90\%$ bound on the number of observations $N$ assumed from a Poisson distribution. We estimate the sensitive volume using a Monte Carlo volume estimation.

The simulated population is assumed to be distributed uniformly in volume and isotropically in orientation and sky position. The injections span fixed chirp masses $\mathcal{M}\in[0.1,0.2,0.35,0.5,0.7,1.1,1.2]~\m$, and  $\tilde{\Lambda}\in[0,1e2,2e3,5e3,6.7e3,9e3,3e4,7e5]$. Injections incompatible with the EOS were excluded, such that a chirp mass of 1.2 would not be paired with a tidal deformability of $7e5$. Since the search was split into 26 sub searches, we apply a trials factor to the most statistically significant candidate.

The rates are presented in Figure~\ref{fig:merger-rate}, where no variation in search sensitivity is observed across tidal deformability. The combined O4a+O4b sensitivity alone has a $1.8$ factor improvement over previous O1-O4a constraints. Incorporating single detector triggers improves the sensitive volume time by $\sim20\%$ for both observing runs.

\section{Constraints on the Primordial Black Hole DM Fraction}

\begin{figure}
    \includegraphics[width=\linewidth]{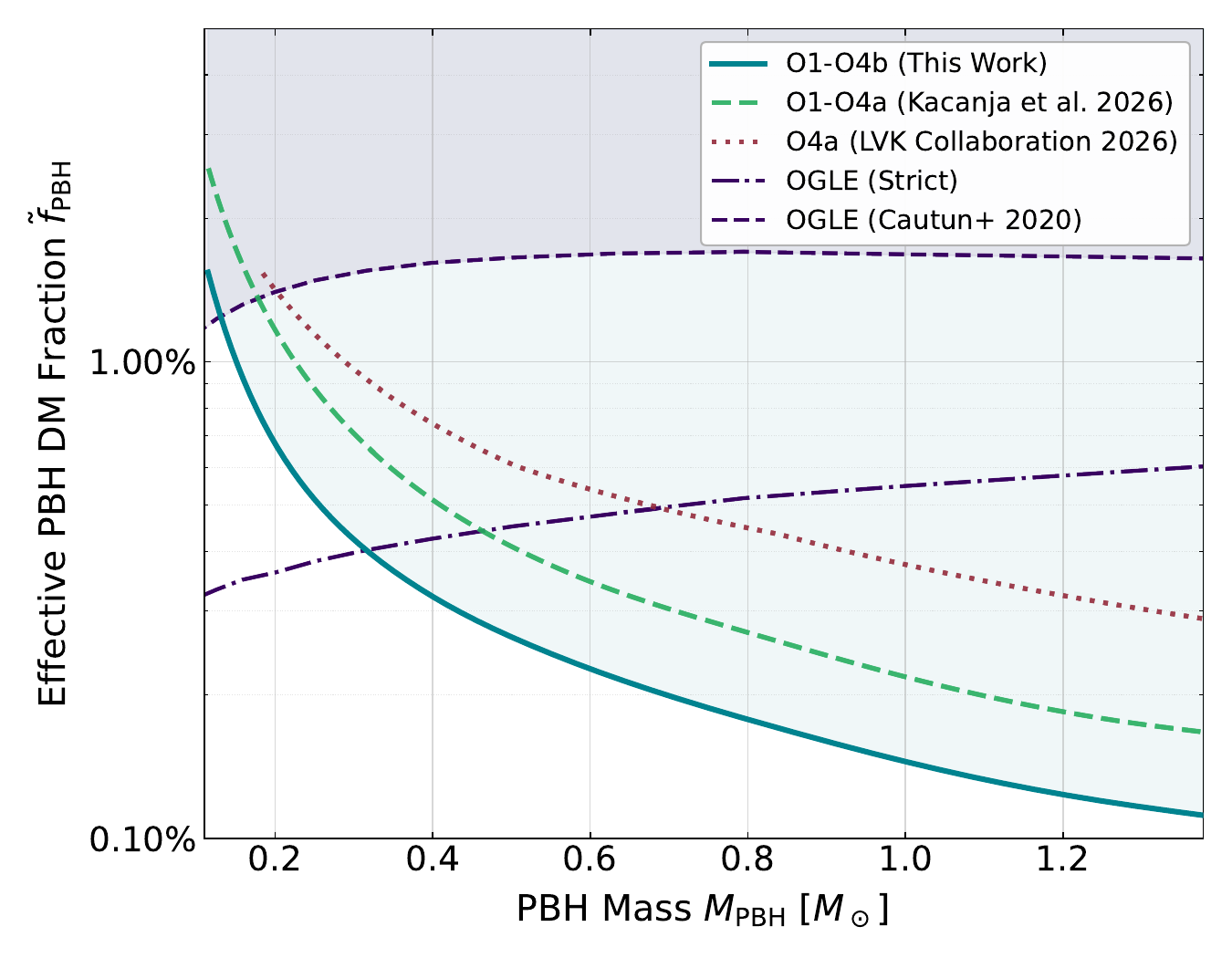}
    \caption{Constraints on the effective PBH dark matter fraction, $\tilde{f}_{\rm PBH}$, as a function of PBH mass $M_{\rm PBH}$, assuming a monochromatic mass distribution and the early-Universe binary formation model of \citet{2017PhRvD..96l3523A} and \citet{2018ApJ...864...61C}. The teal curve shows the cumulative constraints obtained from the combined O1-O4b search presented in this work. Previous GW constraints from the LVK O4a SSM search \citep{lvko4aSSM} and our prior O1-O4a search\citep{Kacanja:2026byy} are shown in dotted dark red and dashed green line respectively. The purple dashed curve and shaded region show the relaxed OGLE microlensing constraints, which make no assumptions about the origin of the observed microlensing events, derived using the halo model of \citet{2020MNRAS.494.4291C}. The purple dash-dotted curve shows stringent OGLE constraints obtained under the assumption that all observed microlensing events arise from known stellar populations \citep{Mroz:2024mse}. The updated GW constraints provide complementary information to the relaxed OGLE limits with a 1.3-13 factor lower for masses $M>0.15$.
}
    \label{fig:dmfrac}
\end{figure} 

Our upper limits on the merger rate of SSM compact binaries can be interpreted as constraints on the fraction of DM $f_{\text{PBH}}$ composed of PBHs. While the measured merger-rate limits are largely model independent, converting them into bounds on the PBH abundance requires assumptions regarding the PBH binary formation channel. We adopt the early-Universe binary formation scenario of \citet{2017PhRvD..96l3523A,2018ApJ...864...61C} as our fiducial model.

For a given PBH mass distribution $P(m)$, the expected merger rate $ R(\tilde{f}_{\rm PBH}, m_1, m_2)$, as a function of the effective fraction of DM $\tilde{f}_\text{PBH}$ and component masses $m_1$ and $m_2$, is given by

\begin{equation}
\begin{split}
    R(\tilde{f}_{\rm PBH}, m_1, m_2,t) = {} & 3 \times 10^6 \tilde{f}_{\rm PBH}^2 (0.7\tilde{f}_{\rm PBH}^2 + \sigma_{eq}^2)^{-21/74} \\
    & \times (m_1 m_2)^{3/37} (m_1 + m_2)^{36/37} \\
    & \times \min \left( \frac{P(m_1)}{m_1}, \frac{P(m_2)}{m_2} \right) \\
    & \times \left( \frac{P(m_1)}{m_1} + \frac{P(m_2)}{m_2} \right)\,\times \left(\frac{t}{t_{0}}\right)^\frac{-34}{37},
\end{split}
\end{equation}

where the age of the universe is given by $t_0$, and the cosmic time in years by $t$. We set $\sigma_{eq} = 0.005$ to parameterize the variance of the dark matter density field evaluated at the epoch of matter-radiation equality \citep{2017PhRvD..96l3523A}. The $\tilde{f}_{\text{PBH}}=S^\frac{37}{53}f_{\text{PBH}}$ and $S$ is the suppression factor that accounts for the disruption of primordial binaries after formation \citep{Raidal:2018bbj,Hutsi:2020sol}. For small PBH abundances ($f_{\text{PBH}}\lesssim0.1\%$), this suppression has only a modest effect on the inferred limits \citep{Raidal:2018bbj,Jedamzik:2020ypm,Hutsi:2020sol}.

The increased sensitivity and observing time of the combined O1-O4b rates yield the most stringent GW constraints to date on the PBH dark matter fraction in the SSM range. At a PBH mass of $0.8~\m$, we constrain the effective dark matter fraction to $\tilde{f}_{\rm PBH}<0.18\%$ at 90\% confidence, representing an improvement of approximately $1.5$ over previous SSM tidal search \citep{Kacanja:2026byy}. These limits further reduce the allowed parameter space for PBHs as dark matter under the assumed binary formation scenario.

We compare our constraints with the latest limits from the OGLE microlensing survey \citep{Mroz:2024mse}. The more stringent OGLE limits assume that all observed microlensing events originate from known stellar populations. In contrast, the more conservative limits make no assumptions regarding the origin of the observed events. Under the binary formation model adopted here, our GW constraints improve upon the relaxed OGLE limits by approximately $\sim9$ for $M_{\rm PBH}>0.8~\m$ and exceed the stringent OGLE limits above $\sim 0.3~\m$, demonstrating the complementary sensitivity of GW searches. The agreement between GW and microlensing constraints provides some consistency between independent probes of DM abundance, although the two constraints rely on different astrophysical assumptions.

\section{Discussions and Conclusion} \label{sec:conclusion}

A confirmed detection of an SSM binary would challenge standard compact-object formation scenarios, which do not predict compact remnants with masses below $\sim 1~\m$, and would motivate the investigation of alternative formation channels. If the SSM object is identified as a neutron star, such a system could be consistent with fragmentation processes proposed to produce multiple ultra-light binaries \citep{Metzger_2024}. Environmental effects of the disk make this scenario challenging to observe \citep{Lerner:2025dkd}, and any observed SSM neutron star may have to form in an isolated scenario, which is currently unexplained by current formation channels. Conversely, the absence of measurable tidal deformations could disfavor a neutron star interpretation and motivate the consideration of SSM black hole of primordial origin as a possibility \citep{darkmatterreview,PBHreview}.

In this work we conducted a dedicated SSM binary search using both the O4b and O4a data from the Advanced LIGO detectors. This search incorporates tidal deformability consistent with the EOS constrained by the GW170817 observation for neutron stars down to $0.25~\m$. We searched for objects with masses between 0.1 and $2~\m$ and effective tidal deformabilities from 0 to $7e5$. No statistically significant candidates were identified. Combining both the O4b and O4a dataset, we constrain the merger rate for SSM objects, improving the O1-O4a constraints by a factor of 2.2. For component masses $M\ge0.1~\m$, we place constraints on the fraction of dark matter comprised of PBHs using the O4a+O4b dataset, providing complementary information to existing microlensing constraints \citep{Mroz:2024mse}.

The increased sensitivity achieved during O4b, together with the larger amount of available data, further demonstrates the importance of continuing dedicated searches for rare SSM systems. The coincident observing time was comparable between O4a and O4b, while the total single-detector observing time increased from approximately $90$ days in O4a to $113$ days in O4b. The additional observing time, combined with improved detector sensitivity, increases the surveyed spacetime volume by $40\%$ in O4b relative to the coincident analysis alone and $20\%$ in O4a. Continued improvements in detector sensitivity, together with dedicated searches that incorporate the distinctive tidal signatures of neutron stars, will be important for identifying these systems.

Recent low-significance GW candidates, such as S250818k and S251112cm, highlight the continued interest in SSM compact objects and the importance of coordinated multi-messenger observations \citep{S250818k,ZTF25abjmnps,2022NatAs...6.1444D,Yuan:2024yyo,ligosubsolar,nitzsubsolar1,nitzsubsolar2,2025ApJ...984...61K,Hall:2026gov,S251112cmgcn,Vieira:2026eof,Haque:2026yum}. The astrophysical nature of these candidates remains unresolved. For S250818k, the optical transient ZTF25abjmnps was discovered during the same space and localization region as the trigger \citep{ZTF25abjmnps,S250818k}, however, optical counterpart was resolved to be a supernova, and the significance of the GW candidate still remains uncertain. No convincing electromagnetic counterpart has been identified for S251112cm, which has motivated a PBH interpretation \citep{S251112cmgcn,Vieira:2026eof,Haque:2026yum}. However, like S250818k, the low significance of the event may be consistent with detector noise. As GW detectors continue to improve in sensitivity, future observations will provide greater sensitivity to weak SSM signals and may help distinguish astrophysical signals from terrestrial noise. The release of the O4c data also provides an opportunity to search for previously undetected SSM BNSs, particularly systems with large tidal deformabilities that may be more readily identified using dedicated tidal searches, while allowing the recovery of these two recent SSM candidates in offline search pipelines.

If SSM systems are scarce or predominantly located at distances far beyond the reach of current detectors, next-generation observatories will be necessary to probe these larger volumes. Detectors such as Cosmic Explorer and Einstein Telescope are expected to extend the observable volume by orders of magnitude relative to current sensitivities \citep{2010CQGra..27a5003H,2010CQGra..27s4002P,2019BAAS...51g..35R,Evans:2023euw}. Cosmic Explorer may be able to probe systems at redshifts up to $\sim 100$, where early PBHs could form and potentially observe these objects \citep{greenPBHreview}.

This work underscores the new generation of unique binary searches and the need for efficient search methodologies. The long duration of SSM objects, combined with tidal effects imprinted on the waveform, drives the size of the template bank. As new, more efficient methods are introduced, and advancements to the FIR framework are made, extending the tidal range of the EOS to the full physically allowed space ($\mathcal{O}(10^8)$ for $0.1~\m$) may be achievable. Furthermore, exotic models such as the formation of SSM BNSs through disk fragmentation may require a more sophisticated GW model that accounts for environmental effects and eccentric obits \citep{Lerner:2025dkd}; this should also increase the cost of conducting a blind search.

\begin{acknowledgments} 

KK, KS, and AHN acknowledge support from NSF grant PHY-2309240. KK and AHN acknowledge the support from Syracuse University for providing the computational resources through the OrangeGrid High Throughput Computing (HTC) cluster supported by the NSF award ACI-1341006. KK acknowledges the support from the Open Science Grid (OSG) for providing additional computational resources to perform the search, which is supported by the National Science Foundation awards 2030508 and 2323298. Additionally, KK would like to express her gratitude to Rahul Dhurkunde, Ananya Bandopadhyay, Yifan Wang, and Labani Roy for all the discussions and feedback received. KK would like to thank Bradley J. Kavanagh for his efforts in maintaining and organizing the \textsc{PBHBounds} GitHub repository consisting of previous PBH constraints \citep{PBHbounds}, which greatly improved the comparative analysis in this work.

This research has made use of data or software obtained from the Gravitational Wave Open Science Center \citep{gwosc}, a service of the LIGO Scientific Collaboration, the Virgo Collaboration, and KAGRA. This material is based upon work supported by NSF's LIGO Laboratory which is a major facility fully funded by the National Science Foundation, as well as the Science and Technology Facilities Council (STFC) of the United Kingdom, the Max-Planck-Society (MPS), and the State of Niedersachsen/Germany for support of the construction of Advanced LIGO and construction and operation of the GEO600 detector. Additional support for Advanced LIGO was provided by the Australian Research Council. Virgo is funded, through the European Gravitational Observatory (EGO), by the French Centre National de Recherche Scientifique (CNRS), the Italian Istituto Nazionale di Fisica Nucleare (INFN) and the Dutch Nikhef, with contributions by institutions from Belgium, Germany, Greece, Hungary, Ireland, Japan, Monaco, Poland, Portugal, Spain. KAGRA is supported by Ministry of Education, Culture, Sports, Science and Technology (MEXT), Japan Society for the Promotion of Science (JSPS) in Japan; National Research Foundation (NRF) and Ministry of Science and ICT (MSIT) in Korea; Academia Sinica (AS) and National Science and Technology Council (NSTC) in Taiwan

\end{acknowledgments}

\section*{Data Availability}
All necessary files used in this work are publicly available in our GitHub repository \citep{github_o4b_sub}.

\bibliography{references}

\end{document}